\documentstyle[sprocl,psfig]{article}

\def\Journal#1#2#3#4{{#1} {\bf #2}, #3 (#4)}

\def\NPB{{\em Nucl. Phys.} B}
\def\PLB{{\em Phys. Lett.}  B}
\def\PRL{\em Phys. Rev. Lett.}
\def\PRD{{\em Phys. Rev.} D}
\def\ZPC{{\em Z. Phys.} C}
\def\PTP{\em Prog.~Theor.~Phys.}
\def\ibid{\it ibid.}

\def\be{\begin{equation}}
\def\ee{\end{equation}}
\def\bea{\begin{eqnarray}}
\def\eea{\end{eqnarray}}

\begin{document}

\newcommand{\gsim}{\mbox{ \raisebox{-1.0ex}{$\stackrel{\textstyle >}
{\textstyle \sim}$ }}}
\newcommand{\lsim}{\mbox{ \raisebox{-1.0ex}{$\stackrel{\textstyle <}
{\textstyle \sim}$ }}}

\begin{flushright}
  \begin{tabular}[t]{l} 
  KEK-TH-606\\
  November 1998
 \end{tabular}
 \end{flushright}
\vspace*{0.5cm}

\title{SUSY Phenomenology
\footnote{
Lecture given at the IVth International Workshop on Particle Physics
Phenomenology, June 18 -21, 1998, Kaohsiung, Taiwan. 
}
}

\author{Yasuhiro Okada}

\address{KEK, Oho 1-1, Tsukuba, 305-0801 Japan } 




\maketitle\abstracts{Three topics on phenomenology in supersymmetric
models are reviewed, namely, the Higgs sector in the supersymmetric model,
B and K meson decay in the supergravity model and lepton flavor violation
processes in supersymmetric grand unified theory.
}

\section{Introduction}
Supersymmetry (SUSY) is a symmetry between fermions and bosons
and its theoretical development started
from the formulation of SUSY invariant Lagrangian in four dimensional 
field theory in early '70s.\cite{WZ} Subsequently, SUSY gauge theory, 
supergravity\cite{FFN} and string theory with space-time supersymmetry 
(supersrting)\cite{GSO} were formulated in '70s.

Phenomenological application of SUSY theory has developed since 
early '80s in connection with naturalness problem in the Standard
Model (SM). One interesting possibility beyond the SM is an idea 
that various interactions are unified at very high energy scale.
Examples are grand unified theory (GUT) and string unification. 
In such theories the cutoff energy scale for 
the SM is considered to be close to the Planck scale ($\sim 10^{19}$ GeV)
and the quadratic divergence of the Higgs field in the SM is very
problematic because very precise fine tuning between the bare mass term 
and radiative correction is necessary to ensure that the weak scale 
is smaller by  $10^{16}$ orders than the Planck scale.
In SUSY theories this problem can be avoided due to cancellation of 
the quadratic divergence in scaler mass renormalization. From this 
point of view SUSY extensions of the SM and the GUT have been
extensively studied for more than 15 years.\cite{nilles}

In the experimental side, the SM has been very successful.             
LEP and SLAC experiments at the Z pole showed quantitative 
agreement of various observables with SM predictions. 
It was also pointed out that three gauge coupling constants determined
precisely there are consistent with the SUSY SU(5) GUT assumption
although non-SUSY version of the simple SU(5)GUT fails to reproduce
the coupling unification.\cite{SUSYCU} Since then it has become more and more 
important to know how one can explore SUSY in the present and
future high energy experiments.

In this lecture after a brief introduction to the minimal supersymmetric
standard model (MSSM) we discuss three topics on SUSY phenomenology,
namely the Higgs sector in SUSY models, B and K meson decay in 
the supergravity model, and lepton flavor violation (LFV) in SUSY GUT.

\section{Minimal Supersymmetric Standard Model}
\subsection{MSSM Lagrangian}
Minimal supersymmetric standard model (MSSM) is a minimal SUSY extension
of the SM. In this model we introduce a SUSY partner for each particle
in the SM. For left- and right-handed quark and lepton complex 
scalar fields, squark and slepton, are introduced. The superpartner
of gauge boson is gauge fermion (or gaugino) and that of the Higgs
field is called higgsino.  The superpartners of gluon and SU(2) and
U(1) gauge fermions are gluino, wino and bino, respectively. After
the electroweak symmetry breaking the wino, bino and higgsino mix 
each other and form two charged Dirac fermions called chargino and
four neutral Majorana fermions called neutralino. As for the Higgs 
sector the SUSY model contains at least two Higgs doublet fields.
This is because different Higgs doublet fields should be introduced
for mass terms for the up-type quarks and those for the down-type
quarks and charged leptons in order to write appropriate Yukawa 
couplings without conflict with SUSY. Two Higgs doublets are also 
required from the gauge anomaly cancellation because the higgsino
fields generate additional contributions to the anomaly cancellation
condition. The particle content of the MSSM is therefore summarized
as two Higgs doublet SM with bosonic superpartners (squarks and sleptons)
and fermionic superpartners (gluino, charginos and neutralinos).

The MSSM Lagrangian consists of two parts, SUSY invariant part and
soft SUSY breaking terms. General rule of writing SUSY invariant 
Lagrangian is well known.\cite{WB} There are two kinds of supermultiplets,
{\it i.e.} gauge supermultiplet which contains gauge fields ($A_\mu$) and
gauge fermion ($\lambda$) and chiral supermultiplet which consists
of a pair of a complex scalar field ($\phi$) and a left-handed Weyl
fermion ($\psi$). The SUSY invariant interactions are classified 
essentially to two kinds, {\it i.e.} gauge interactions and 
superpotential interactions. 

The gauge interaction is completely determined once we fix gauge 
group and representation of each chiral supermultiplet. In addition
to ordinary gauge interactions between matter fields and gauge fields
through covariant derivatives there are two additional interactions
specified by the gauge coupling constant. One is gaugino-fermion-scalar
coupling and the other is scalar four-point coupling called D term.
This kinds of relation between bosonic and fermionic interactions
are important feature of SUSY theory.

The superpotential interaction is a set of scalar potential and fermion 
mass terms or Yukawa type interactions. This is specified by a scalar
function called superpotential. In the MSSM we take
\begin{eqnarray}
W & = & (y_u)_{ij}H_2 U_i^c Q_j +(y_d)_{ij}H_1 D_i^c Q_j
\nonumber\\
  & & +(y_e)_{ij}H_1 E_i L_j +\mu H_1 H_2.
\end{eqnarray}
The SU(3)$\times$ SU(2)$\times$ U(1) quantum numbers for each superfield 
are given by 
$ Q({\bf 3},{\bf 2},\frac{1}{6}),$ 
$U^c({\bf \bar{3}},{\bf 1},-\frac{2}{3}),$
$D^c({\bf \bar{3}},{\bf 1},\frac{1}{3}),$
$L({\bf 1},{\bf 2},-\frac{1}{2}),$
$E({\bf 1},{\bf 1},1),$
$H_1({\bf 1},{\bf 2},-\frac{1}{2}),$
$H_2({\bf 1},{\bf 2},\frac{1}{2})$ 
and $i, j$ are generation indices. If we only demand gauge
invariance we can also include the following coupling in the 
superpotential.  
\begin{equation}
W_{\not R } =\lambda_1 QLD^c + \lambda_2 LEL + \lambda_3
U^c D^c D^c.
\end{equation}
These couplings in general violate the lepton number and/or baryon number
conservation, therefore unless some of these coupling constants are 
extremely small too fast proton decay is induced. The most popular way to
forbid the above terms in the superpotential is to require conservation 
of a multicative quantum number called R parity which is assigned to be
+1 for ordinary particles and -1 for superpartners. The R parity 
conservation has an important phenomenological consequence. The 
lightest SUSY particle (LSP) becomes stable. Since LSP should be 
a neutral particle from cosmological consideration LSP becomes 
a natural candidate of dark matter of the Universe. The existence 
of LSP is also important in the direct SUSY search experiment because 
SUSY particles should be pair-produced and decay to final states 
including missing energy.
\subsection{Soft SUSY Breaking Terms}
The soft SUSY breaking terms are defined as those terms in the Lagrangian
which violate SUSY invariance but do not induce quadratic divergence.
In this way the naturalness is maintained as long as the SUSY 
breaking scale is close to the weak scale. The general form of the 
soft SUSY breaking terms are classified.\cite{GG} These terms are 
gaugino mass terms, scalar mass terms and triple and quadratic 
couplings among scalar components of 
chiral supermultiplets (or anti-chiral supermultiplets).
\begin{eqnarray}
L_{SUSY Breaking} & = & -\frac{1}{2}\Sigma M_\alpha \bar{\lambda_\alpha}
{\lambda_\alpha}-\Sigma m_i^2|\phi_i|^2
\nonumber\\
  & & +(A_u H_2 {\tilde u}^c {\tilde q} 
      +A_d H_1 {\tilde d}^c {\tilde q}
      +A_eH_1 {\tilde e} {\tilde l}) + h.c.
\nonumber\\
    & &  + B \mu H_1 H_2 + h.c..
\end{eqnarray}   
Here we suppress the generation indices and neglect the
generation mixing in the scalar mass terms.  
The role of the SUSY breaking terms is basically to provide mass terms 
for SUSY partners so that SUSY particles can be heavier than the
ordinary particles.

The origin of these soft SUSY breaking terms can be thought of 
the remnant of spontaneous SUSY breaking. If we consider the MSSM 
Lagrangian as a low energy effective theory of some unification 
theory valid at very high energy scale, it is natural to take gravity 
interaction into consideration. Once gravity is included SUSY becomes
local invariance, {\it i.e.} supergravity theory. Because local invariance 
is necessary for consistency of the theory the only way to break
the symmetry is the spontaneous one. If we want to
construct a model based on spontaneous SUSY breaking at the tree level
in global SUSY theory, there exists a tree level mass relation
which severely constrains the model building.\cite{massrel}
The phenomenologically viable way to generate the soft SUSY breaking 
terms in the MSSM Lagrangian is therefore to prepare a separate sector 
where the spontaneous SUSY breaking occurs and to couple this sector 
with the MSSM through some weak interaction such as gravity or loop 
effects. As a result appropriate soft SUSY breaking terms are generated 
in the MSSM Lagrangian.

There are important phenomenological constraints on the SUSY
breaking terms from flavor and CP violation physics. 
The SUSY breaking terms for squarks and sleptons masses can 
induce too large flavor changing neutral current
(FCNC), LFV and neutron and electron electric dipole moments (EDM).
This is because squark and slepton mass matrices can be new sources of
flavor mixings and CP violation which are not related to the 
Cabibbo-Kobayashi-Maskawa (CKM) matrix. These flavor and CP 
problems provide us some clue to consider possible mechanism 
of SUSY breaking.  

For example, if we assume that the SUSY contribution to the  
$K^0 - \bar{K}^0$ mixing is suppressed because of the cancellation
among the squark contributions of different generations, the squarks
with the same $SU(3)\times SU(2)\times U(1)$ quantum numbers have 
to be highly degenerate in masses. When the squark mixing angle is 
in a similar magnitude to the Cabibbo angle the requirement on 
degeneracy becomes as
\begin{equation}
\frac{\Delta m_{\tilde{q}}^2}{ m_{\tilde{q}}^2}\lsim 10^{-2}
\left(\frac{ m_{\tilde{q}}}{100 GeV}\right)
\end{equation}
for at least the first and second generation squarks.
Similarly, the $\mu^+ \rightarrow e^+ \gamma$ process
puts a strong constraint on the flavor off-diagonal terms
for slepton mass matrices which is roughly given by
\begin{equation}
\frac{\Delta m_{\tilde{\mu}\tilde{e}}^2}{ m_{\tilde{l}}^2}\lsim 10^{-3}
\left(\frac{ m_{\tilde{l}}}{100 GeV}\right)^2 .
\end{equation}

There are several attempts to overcome this problem. One popular 
assumption is that SUSY breaking mass terms are generated through
the coupling of supergravity in a flavor-blind way. In the minimal 
supergravity model all the scalar fields are assumed to be universal
at the Planck scale and therefore there are no FCNC and LFV at this scale. 
Since physical squark and slepton masses are defined at the weak scale 
we have to evaluate these mass matrices using the renormalization group 
equations (RGE) with initial conditions at the Planck scale. As discussed 
later the squarks for the first and second 
generations can be highly degenerate at the weak scale in this model and the 
$K^0 - \bar{K}^0$ mixing from the SUSY loop can be suppressed. 
Another idea is so called gauge mediated SUSY breaking 
model.\cite{gaugemed} In this case
SUSY is assumed to be dynamically broken in a separate sector and
the SUSY breaking effect is transmitted to the MSSM sector by loop 
effects of extra vector-like quarks and leptons. Since this  transmission
is based on gauge interaction the squarks and sleptons with the same gauge
quantum numbers receive the same magnitude of the SUSY breaking effect. 
Thus the degeneracy necessary to avoid the FCNC problem is
guaranteed in this scenario. A general consequence of this scenario
is that unlike the gravity-mediated case gravitino becomes LSP.
The strategies of SUSY particle search is also quite different in this  
scenario which is performed in recent collider experiments.\cite{exgm}  

\section{Higgs Sector in the MSSM} 

\subsection{The SM Higgs sector}
Study of the Higgs sector is not only important to establish the 
SM but also crucial to search for physics beyond the SM. In this
respect the mass of the Higgs boson itself is the most important
information. For example, a heavy Higgs boson suggests that the
dynamics of the electroweak symmetry breaking is governed by strong
interaction. If we assume that fundamental
interactions are described by perturbation theory up to
the Planck scale or a scale close to it, the Higgs boson is 
expected to exist below 200 GeV. GUT and SUSY extension of the SM
are examples of the latter case.

This situation is more clearly shown with a help of the SM RGE.  
In the minimal SM the Higgs-boson mass ($m_h$) and the Higgs-boson
self-coupling constant ($\lambda$) from the Higgs potential
($V_{Higgs}=m^2 |\Phi|^2 + \lambda|\Phi|^4$,
$(m^2\le0)$) are related by $m_h^2 = 2\lambda \upsilon^2$,
where $\upsilon (= 246$ GeV) is the vacuum expectation value.
Neglecting Yukawa coupling constants except for the one for
the top quark, $y_t$, the RGE for the running
self-coupling constant $\lambda$ at the one-loop level is written as
\begin{equation}
\frac{d\lambda}{dt} = \frac{1}{(4\pi)^2}
\{24\lambda^2 + 12y_t^2\lambda - 6y_t^4-12A\lambda+6B\},
\label{rge}
\end{equation}
where $A = \frac{1}{4}g_1^2 + \frac{3}{4}g_2^2$,
$B = \frac{1}{16}g_1^4 + \frac{1}{8}g_1^2g_2^2 + \frac{3}{16}g_2^4$
for $U(1)$ and $SU(2)$ gauge coupling constants $g_1$ and $g_2$
respectively and $t=\ln{\mu}$ ($\mu$ is the renormalization scale).
Using the gauge coupling constants and the top Yukawa
coupling constant at the electroweak scale we can draw the flow of the 
running self-coupling constant for each value of $m_h$ as shown in 
Figure 1.
%
\begin{figure}
\begin{center}
\mbox{\psfig{figure=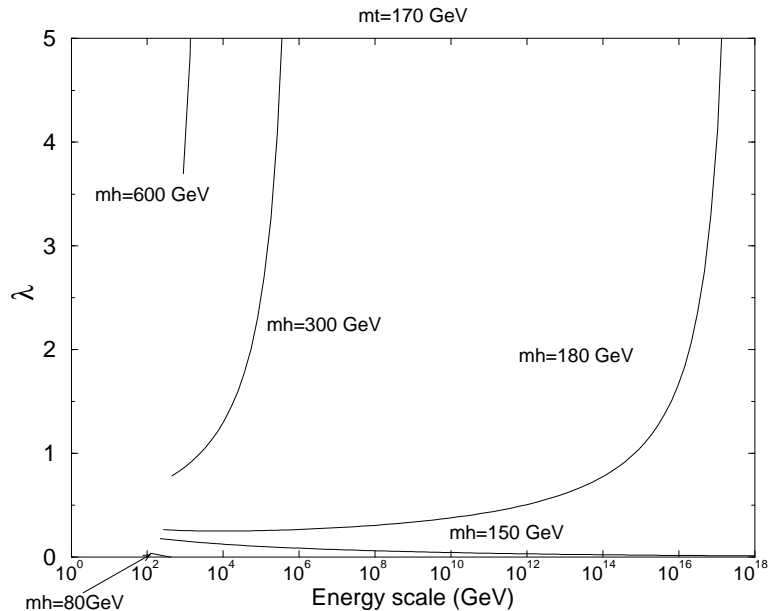,width=4in,angle=-90}}
\end{center}
\caption{The flow of the Higgs self-coupling constant ($\lambda$)
in the minimal SM for the several values of the Higgs
boson masses. The top quark mass is assumed to be 170 GeV.
\label{fig:fig1}}
\end{figure}

 From this figure it is clear that possible scenarios for new
 physics are different for
 two cases, {\it i.e.}\/ (i) $m_h \gg m_{top}$ and
(ii) $m_h \lsim m_{top}$.
 In case (i) the coupling constant becomes very large at a relatively
 low energy scale, and therefore new physics is indicated well
 below the Planck scale.  On the other hand,
 the theory can be
 weakly coupled up to approximately the Planck scale in case (ii),
 which is consistent
 with the idea of grand unification.  Of course the flow
of the coupling
constants is different if we change the particle content
in the low-energy
theory, but the upper bound on the Higgs mass is believed
to be about 200 GeV for most GUT models.
This is also true for the SUSY GUT.
As for the MSSM, however, a stronger bound on the Higgs mass
is obtained
independently of the assumption of grand unification.

\subsection{The Higgs Sector in the MSSM}
The most important feature in the MSSM Higgs sector is
that the Higgs-self-coupling
constant at the tree level is
completely determined by the $SU(2)$ and $U(1)$ gauge coupling
constants. After electroweak symmetry breaking, the physical
Higgs states include two CP-even Higgs bosons ($h, H$),
one CP-odd Higgs boson ($A$) and one pair of charged Higgs
bosons ($H^\pm$) where we denote by $h$
and $H$ the lighter and heavier Higgs bosons, respectively.  
Although at the tree level the
upper bound on the lightest CP-even Higgs boson mass is given
by the $Z^0$ mass, the radiative corrections
can raise this bound.\cite{OYY} The Higgs potential is given by

\begin{eqnarray}
V_{Higgs} & = & m^2_1|H_1|^2 + m_2^2|H_2|^2 - m_3^2
   (H_{1}\cdot H_{2}+\bar{H_{1}}\cdot\bar{ H_{2}})
\nonumber\\
          & & +\frac{g_2^2}{8}(\bar{H_1}\tau^aH_1
                + \bar{H_2}\tau^aH_2)^2
                + \frac{g_1^2}{8}(|H_1|^2 - |H_2|^2)^2
\nonumber\\
          & & + \Delta V,
\end{eqnarray}
where $\Delta V$ represents the contribution from one-loop diagrams.
Since the loop correction due to the top quark and its
superpartner, the stop squark, is proportional to the fourth
power of the top Yukawa coupling constant the
Higgs self-coupling constant is no longer determined only by
the gauge coupling constants.
The upper bound on the lightest CP-even Higgs mass ($m_h$)
can significantly increase for a reasonable choice of the
top-quark and stop-squark masses. Taking account of only 
one-loop effects of  top-quark and stop-squark, the upper
bound is given by
\begin{equation}
m_h^2 \leq m_Z^2 \cos^2{2\beta} +\frac{6}{(2\pi)^2}
\frac{m_t^4}{v^2}\ln{\frac{m_{stop}^2}{m_t^2}},
\label{mass}
\end{equation}
where $\beta$ is the angle determined by  the ratio of two Higgs 
vacuum expectation values ($\tan\beta = \frac{<H_2^0>}{<H_1^0>})$.
Figure 2 shows the upper bound on $m_h$ as a function of top-quark mass
for several choices of the stop mass and $\tan\beta$.
\begin{figure}
\begin{center}
\mbox{\psfig{figure=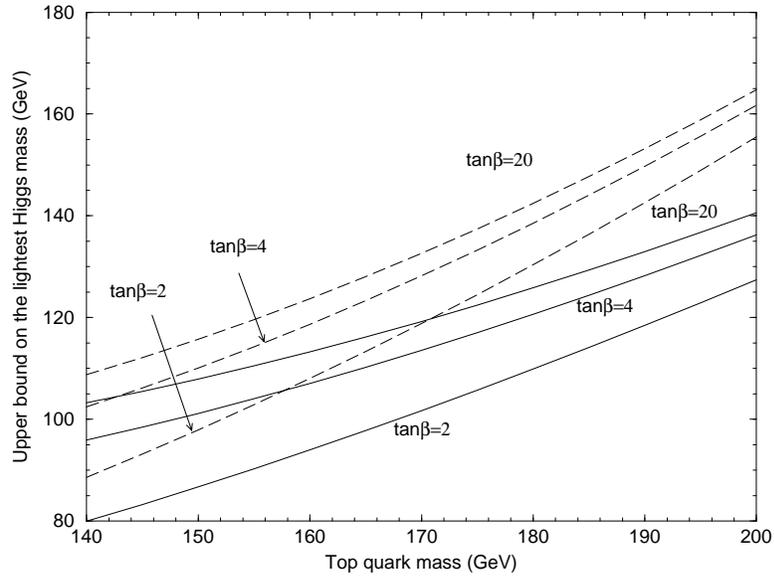,width=4in,angle=-90}}
\end{center}
\caption{The upper bound on the lightest CP-even Higgs mass in the MSSM
as a function of the top quark mass for various $\tan{\beta}$ and
two large stop mass scales. The solid (dashed) line corresponds
to $m_{stop}$=1 (10) TeV without left-right mixing of two stop states.
These masses are calculated by the method with the renormalization
group equation.\protect\cite{OYYRGE}
\label{fig:fig2}}
\end{figure}
We can see that, in the MSSM, at least one neutral Higgs-boson should
exist below 130 - 150 GeV depending on the top and stop masses.

The physical reason for the increase of the upper bound
can be easily understood if we take the SUSY scale much larger than
the weak scale.\cite{OYYRGE,BRGE} Suppose, for example,
that all the SUSY particles
as well as physical Higgs states other than the lightest CP-even
Higgs boson exist around 1 TeV. Then the effective theory
which describes physics below 1 TeV scale is just the
minimal SM with one Higgs doublet. On the other hand at the
energy scale above 1 TeV the MSSM Lagrangian should be recovered.
Information of the SUSY theory can be included in the matching
condition between two theories at the 1 TeV scale. Because 
SUSY relation should be a good approximation above 1 TeV     
the Higgs self-coupling constant $\lambda$ in the effective
theory is given by  $\lambda= \frac{1}{8}(g_1^2 + g_2^2) 
\cos^2{2\beta}$ at 1 TeV. The value of $\lambda$ changes
according to the SM  SM RGE below 1 TeV and the Higgs boson mass is
determined by $\lambda$ at the weak scale. Because of the $y_t^4$
term in Eq.(\ref{rge}) $\lambda$ becomes larger toward low energy
and this results in the increase of the Higgs boson mass. Thus
the logarithmic dependence to the stop mass
in Eq. (\ref{mass}) can be understood as a renormalization effect 
between the SUSY scale and the top scale and the increase
of the upper bound can be interpreted as a correction to
the tree-level SUSY relation due to the mass difference
of top quark and stop squark.  

Other Higgs states, namely the $H, A, H^\pm$, are also important
to clarify the structure of the model. Their existence alone is a proof
of new physics beyond the SM, but we may be able to distinguish the
MSSM from a general two-Higgs model through the investigation of
their masses and couplings.  In the MSSM the Higgs sector is
essentially determined by three independent parameters 
if we neglect mass difference and mixing of left- and right-handed stop
states. For the free parameters we can take the mass of the CP-odd Higgs 
boson ($m_A$), $\tan\beta,$ the stop mass ($m_{stop}$).
Note that the stop masses enter through radiative corrections to
the Higgs potential. 
In Figure 3, the masses for the $H, A$, and $ H^\pm$ are shown as a
function of $m_A$ for several choices of $\tan\beta$
and $m_{stop}$=1 TeV.
%
\begin{figure}
\begin{center}
\mbox{\psfig{figure=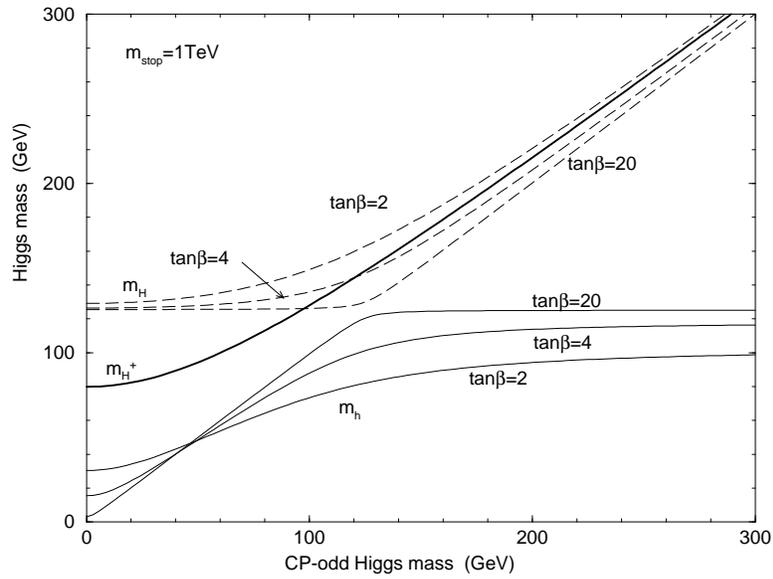,width=4in,angle=-90}}
\end{center}
\caption{The light ($h$), heavy ($H$) CP-even Higgs masses and
the charged
Higgs ($H^\pm$) mass as a function of the CP-odd Higgs ($A$) mass.
The top and stop masses are taken as $m_t$ =~170 GeV and
$m_{stop}$ =~1 TeV.
\label{fig:fig3}}
\end{figure}
We can see that, in the limit of $m_A \rightarrow \infty$, $m_h$
approaches a constant value which corresponds to the upper bound
in Figure 1. Also in this limit the
$H, A$ and $H^\pm$ become degenerate in mass.

The neutral Higgs-boson couplings to gauge bosons and fermions
are determined by $\tan{\beta}$
and the mixing angle $\alpha$
of the two CP-even Higgs particles defined as
\begin{eqnarray}
ReH_1^0 &=& \frac{1}{\sqrt{2}}
(\upsilon\cos\beta - h\sin\alpha + H\cos\alpha)
\nonumber \\
ReH_2^0 &=& \frac{1}{\sqrt{2}}
(\upsilon\sin\beta + h\cos\alpha + H\sin\alpha).
\end{eqnarray}
In the MSSM $\alpha$ is a function of above three independent parameters.
The various coupling constants and therefore properties of Higgs 
bosons are specified by these two angles.
Let us consider the Higgs-boson production in $e^+e^-$ linear
collider experiments. In this case the Higgs-bremsstrahlung process
$e^+e^- \rightarrow Zh$ or $ZH$ and the associated production
$e^+e^- \rightarrow Ah$ or $AH$ play complimentary roles.
Namely $e^+e^- \rightarrow Zh~(ZH)$ is proportional
to $\cos(\beta-\alpha)(\sin(\beta-\alpha))$, and
$e^+e^- \rightarrow Ah~(AH)$ is proportional to
$\sin(\beta-\alpha)(\cos(\beta-\alpha))$,
so at least one of the two processes has a sizable coupling.
It is useful to distinguish the following two cases, namely
(i) $m_A \lsim 150$ GeV, (ii) $m_A \gg 150$ GeV.
In case (i), the two CP-even Higgs bosons can have large mixing, and
therefore the properties of
the neutral Higgs boson can be substantially different from those
of the minimal SM Higgs.  On the other hand, in case (ii), the
lightest CP-even Higgs becomes a SM-like Higgs, and the other
four states, $H, A, H^\pm$ behave as a Higgs doublet orthogonal
to the SM-like Higgs doublet.  In this region,
$\cos(\beta-\alpha)$ approaches unity and $\sin(\beta-\alpha)$
goes to zero so that $e^+e^- \rightarrow Zh$ and
$e^+e^- \rightarrow AH$ are the dominant production processes.

Scenarios for the Higgs physics at a future $e^+e^-$ linear
collider are different for two cases.  In case (i) it is possible
to discover all Higgs states with $\sqrt{s} = 500$ GeV, and the
production cross-section of the lightest Higgs boson may be
quite different from that of the SM so that
it may be clear that the discovered Higgs is not the SM Higgs.
On the other hand, in case (ii), only the lightest Higgs may
be discovered at the earlier stage of the $e^+e^-$ experiment,
and we have to go to a higher energy machine to find the heavier
Higgs bosons. Also, since the properties
of the lightest Higgs boson may be quite similar to those of
the SM Higgs boson we need precision experiments on the production
and decay of the particle
in order to investigate possible deviations from the SM.

\subsection{The Higgs sector in extended versions of the SUSY SM}
Although the MSSM is the most widely studied model, there are
several extensions of the SUSY version of the SM.  If we focus
on the structure of the Higgs sector, the MSSM is special because
the Higgs self-couplings at the tree level
are completely determined by the gauge coupling constants.
It is therefore important to know how the Higgs phenomenology
is different for models other than the MSSM.

A model with a gauge-singlet Higgs field is the simplest
extension.\cite{singlet}  This
model does not destroy the unification of the three gauge coupling
constants since extra particles do not carry the SM quantum
numbers. Moreover, we can include a term $W_\lambda = \lambda NH_1H_2$
in the superpotential where $N$ is a gauge singlet superfield.
Since this term induces $\lambda^2|H_1H_2|^2$ in the Higgs potential,
the tree-level Higgs-boson self-coupling depends on $\lambda$
as well as the gauge coupling constants. Including the one loop
correction the upper bound of the lightest CP-even Higgs-boson mass 
is given by
\begin{equation}
m_h^2 \leq \frac{\lambda^2 v^2}{2}\sin^2{2\beta}+
m_Z^2 \cos^2{2\beta} +\frac{6}{(2\pi)^2}
\frac{m_t^4}{v^2}\ln{\frac{m_{stop}^2}{m_t^2}}.
\label{masssing}
\end{equation}
There is no definite
upper-bound on the lightest CP-even Higgs-boson mass in this model
unless a further assumption on the strength of the coupling $\lambda$
is made.  If we require all dimensionless
coupling constants to remain perturbative up to the GUT scale
we can calculate the upper-bound of the lightest CP-even Higgs-boson
mass.\cite{singletmass} Numerical analysis shows that the upper bound
is about 130 GeV in case that the stop mass is 1 TeV and the top mass 
is 170 to 180 GeV. This is not very much different from the MSSM case 
because for these large top masses we can show that the coupling 
$\lambda$ cannot be so large at the weak scale as a result of the RGE 
analysis.

This result means that the lightest Higgs boson is at least
kinematically accessible at an $e^+e^-$ linear
collider with $\sqrt{s}\sim 300 - 500$ GeV.  This does not necessarily
mean that the lightest Higgs boson is detectable.  In this model
the lightest Higgs boson is composed of one gauge singlet and two
doublets, and if it is singlet-dominated its couplings to the
gauge bosons are significantly reduced, hence its
production cross-section is too small. In such a case the heavier
neutral Higgs bosons may be detectable since these bosons have a
large enough coupling to gauge bosons. In fact we can put an
upper-bound on the mass of the heavier Higgs boson when the
lightest one becomes singlet-dominated.
By quantitative study of the masses and the production
cross-section of the Higgs bosons in this model, we can show that
at least one of the three CP-even Higgs bosons has a
large enough production cross-section in the
$e^+e^- \rightarrow Zh^o_i$ $(i =1, 2, 3)$ process to be detected
at an $e^+e^-$ linear collider with
$\sqrt{s}\sim300 - 500$ GeV.\cite{KOT}  
For this purpose we define the minimal production
cross-section, $\sigma_{min}$, as a function of $\sqrt{s}$ such that
at least one of these three $h^0_i$  has a larger production
cross section than $\sigma_{min}$ irrespective of the
parameters in the Higgs mass matrix.
The $\sigma_{min}$ turns
to be  given by one third of the SM production cross-section
with the Higgs boson mass equal to the upper-bound value.
In Figure 4 we show  $\sigma_{min}$ as a function of $\sqrt{s}$
for $m_{stop} = 1$ TeV .
\begin{figure}
\begin{center}
\mbox{\psfig{figure=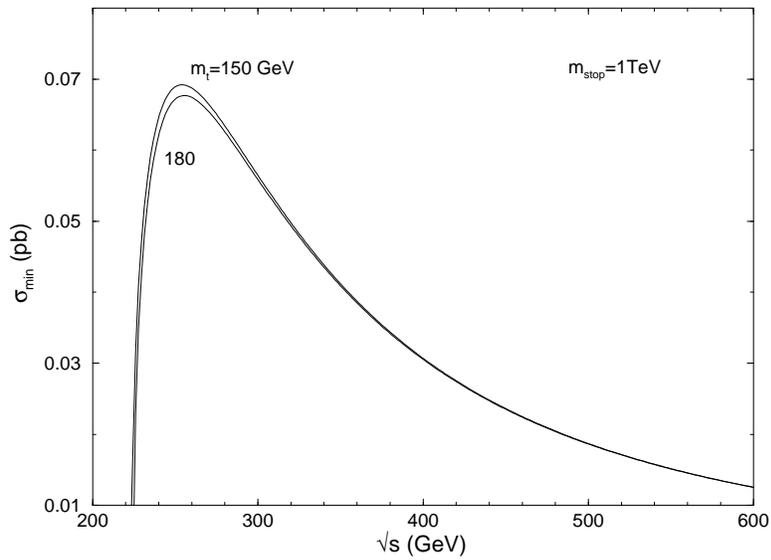,width=4in,angle=-90}}
\end{center}
\caption{Minimal production cross section, $\sigma_{min}$, for
the SUSY SM with a gauge singlet Higgs for the top mass
$m_t$=150 and 180 GeV and $m_{stop}$=1 TeV.
\label{fig:fig4}}
\end{figure}
 From this figure we can conclude that the discovery of 
at least one neutral Higgs boson is guaranteed at an $e^+e^-$
linear collider with an integrated luminosity of 10 fb$^{-1}$.

If we further extend the model we can increase the upper bound 
of the lightest CP even Higgs boson mass. One such possibility 
is to include extra matter fields such as extra families. Then 
loop effect of extra matter fields can raise the upper bound
if Yukawa coupling constants associated to the extra families
are large. If we require that all the dimensionless coupling
constants remain perturbative up to the GUT scale we can restrict
possible content of extra matter fields and also the magnitude
of relevant Yukawa coupling constants. This kind of analysis 
was performed  for models with extra matter multiplets which 
keep the SU(5)gauge coupling unification.\cite{MO} It was shown that the
upper bound can be as large as 180 GeV for 1 TeV SUSY scale
in the model with extra $\bf{10}+\bf{\bar{10}}$ or
$\bf{10}+\bf{\bar{5}}+\bf{\bar{10}}+\bf{5}$ representations
of SU(5) gauge group. 

Another possibility is to consider triplet Higgs fields 
instead of a gauge singlet field.\cite{triplet} A analysis similar to the
model with a gauge singlet Higgs field leads to the maximum
value of the lightest CP-even Higgs boson mass is about
180 GeV if we require that all the dimensionless coupling     
constants remain perturbative below the GUT scale.
 
\subsection{Phenomenology of MSSM Higgs searches}

The Higgs boson in the intermediate mass region such as 
the lightest Higgs boson in the MSSM is a target 
of the future collider experiments both at LHC and $e^+ e^-$ 
linear colliders. In the experiment at LEP II the SM Higgs boson 
is expected to be discovered if its mass is below 105 GeV and
the upgraded Tevatron experiment also cover a similar mass
region. Since the upper bound exceeds the discovery limit of 
LEP II it is very important to know the discovery potential 
of the SUSY Higgs in the LHC experiment. In this mass range the main 
decay mode of the SM Higgs boson is $h\rightarrow b \bar{b}$. 
But because of QCD backgrounds of this mode the LHC Higgs boson search
mainly relies on the two photon mode whose branching ratio is 
$O(10^{-3})$. In the SUSY case its branching ratio can be even 
smaller. However in some parameter space,
especially for large $\tan{\beta}$,
 $H, A, h \rightarrow \tau^+ \tau^-$ modes are useful.
Recent study shows that it is possible
to observe at least one mode of the MSSM Higgs signals in
all parameter space after several years of running at
high luminosity.\cite{LHC} 

On the other hand an $e^+ e^-$ linear collider with 
$\sqrt{s}\sim300 - 500$ GeV
is a suitable place to study the Higgs boson in this 
mass region. Here we can not only discover the Higgs boson
easily but also measure various quantities, {\it i.e.}
production cross sections and branching ratios related
to the Higgs boson.\cite{Janot,Kawagoe,JLC,Hildreth}
These measurements are very important to clarify nature 
of the discovered Higgs boson and distinguish the
SM Higgs boson from Higgs particles associated with
some extensions of the SM like the MSSM.     
%
%
\begin{figure}
\begin{center}
\mbox{\psfig{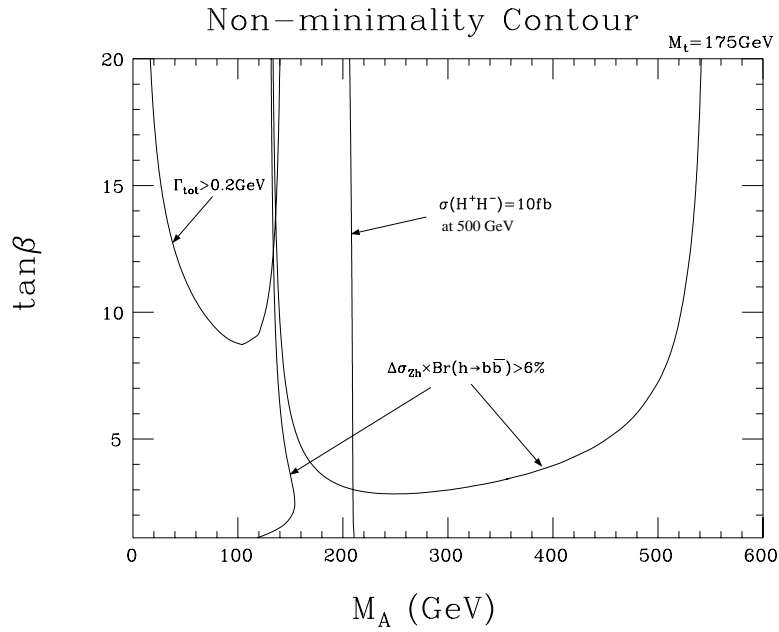}}
\end{center}
\caption{Parameter space in the MSSM where deviation
from the SM is expected to be observable at $e^+e^-$ 
linear-collider. Three lines corresponds to: 
(1) The total width of the lightest CP even Higgs boson mass 
is larger, (2) $\sigma (e^+e^- \rightarrow Z h) 
Br (h\rightarrow b \bar{b})$
deviates from the SM value more than  6\% for $\sqrt{s}=300$ GeV,
(3) $\sigma (e^+e^- \rightarrow H^+H^-) > 10$ fb for 
$\sqrt{s}=500$ GeV. This figure is provided by A. Miyamoto.
\label{fig:fig5}}
\end{figure}
In Figure \ref{fig:fig5} the parameter region of the MSSM
is shown where some deviation from the SM is expected
to be observable at $e^+e^-$ linear-collider. In this figure
we take $m_{stop}=1$ TeV and $m_{t}=175$ GeV. There are three 
lines in this figure which
corresponds to: (1)The total width of the lightest CP even 
Higgs boson mass is larger than 200 MeV which is expected 
to be measurable using leptonic decay mode of Z boson in the 
$e^+e^- \rightarrow Z h$ reaction. (2) The number
of $e^+e^- \rightarrow Z h$, $h \rightarrow b \bar{b}$ events
deviates from the SM prediction more than 6\% for 
$e^+e^-$ linear-collider with $\sqrt{s}=300$ GeV.
(3) The discovery limit of charged Higgs pair production
at $e^+e^-$ linear-collider with $\sqrt{s}=500$ GeV.
We can see that there are sizable deviation
from the SM in a large part of the parameter space 
by measuring the production cross multiplied by the 
$h \rightarrow b \bar{b}$ branching ratio.

Another important information on the MSSM parameter 
is obtained measuring various branching ratio
of the Higgs boson.
We show in particular the ratio of the
following branching ratios,
\begin{equation}
R_{br}\equiv \frac{Br(h \rightarrow c\bar{c}) + Br(h \rightarrow gg)}
{Br(h \rightarrow b\bar{b})},
\end{equation}
is useful to constrain the heavy Higgs mass 
scale if we assume that only one CP-even Higgs boson
is discovered at the first stage of the $e^+e^-$ linear-collider 
experiment \cite{Kamoshita}.

If we assume that the lightest 
CP-even Higgs boson mass ($m_h$) is precisely known we can solve 
for one of the three free parameters $m_A$, $\tan\beta$ and 
$m_{stop}$ in terms of the other parameters. For example, if we solve
for $\tan\beta$ the unknown parameters become $m_A$ and $m_{stop}$.
In Figure 6 the above ratio $R_{br}$ is shown as a function
$m_A$ for different values of $m_{stop}$. We can see that 
$R_{br}$ is almost independent of $m_{stop}$.
%
%
\begin{figure}
\begin{center}
\mbox{\psfig{figure=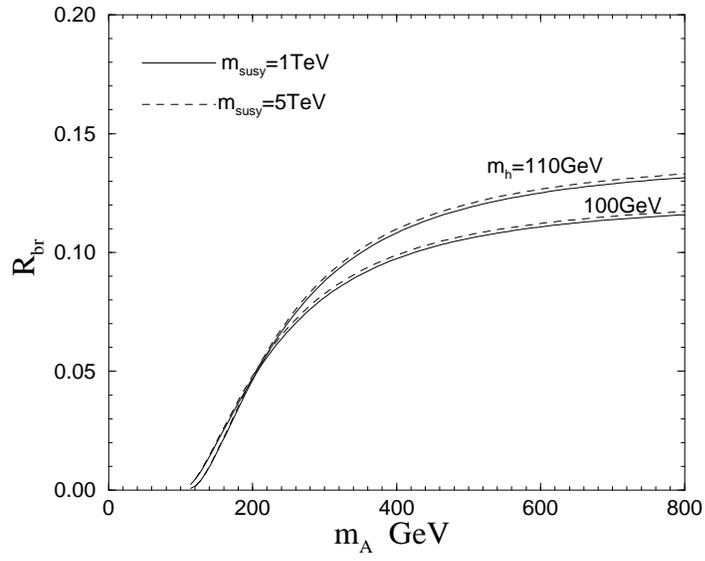,width=4in,angle=0}}
\end{center}
\caption{
$R_{br}\equiv
\protect\frac{(Br(h\to c\bar{c})+Br(h\to gg))}{Br(h\to b\bar{b})}$
as a function of $m_A$ for $m_{susy} =1, 5$ TeV and
$m_h=100, 110$ GeV.\protect\cite{Kamoshita}
The following parameters are used for the calculation of
the branching ratios: $m_t= 170$ GeV, $\bar m_c(m_c)=1.2$ GeV,
$\bar m_b(m_b)=4.2$ GeV, $\alpha_s(m_Z)=0.12$.
\label{fig:fig6}}
\end{figure}

This simple dependence can be explained in the following way.
In the MSSM, each of the two Higgs doublets couples to either
up-type or down-type quarks.  Thus the ratio of the Higgs
couplings to up-type quarks and to down-type quarks is
sensitive to the parameters of the Higgs sector, {\it i.e.} the angles
$\alpha$ and $\beta$.
Since the gluonic width of the Higgs boson is also generated by a
one-loop diagram with an internal top-quark,
the Higgs-gluon-gluon coupling is essentially
proportional to the Higgs-top coupling.
Thus $R_{br}$ is proportional to square of the ratio of
the up-type and down-type Yukawa coupling constants which 
depend on two angles as $(\tan{\alpha} \tan{\beta})^{-2}$.
By examining the neutral Higgs boson mass matrix in the MSSM
at one loop level we can derive approximate 
relation for $R_{br}$ in the MSSM, normalized by $R_{br}$ in the SM as
\begin{equation}
\frac{R_{br}(MSSM)}{R_{br}(SM)}\approx \left(\frac{m_h^2 - m_A^2}
{m_Z^2 + m_A^2}\right)^2
\end{equation}
for $m_A \gg m_h \sim m_Z$. This is actually a very good 
approximation so that $R_{br}$ is almost independent of $m_{stop}$.
  
Measuring this quantity to a good accuracy is therefore 
important for constraining the scale of the heavy
Higgs mass and in choosing the beam energy of 
the $e^+e^-$ linear-collider at th second stage 
when only one Higgs boson is discovered at the first stage.
Note that $R_{br}$ approaches the SM value rather slowly in the
large $m_A$ limit. We can see that $R_{br}$ is reduced by 20$\%$
even for $m_A = 400$ GeV. By simulation study for 
$e^+e^-$ linear collider experiments it is shown that  
the sum of the charm and gluonic branching ratios 
can be determined reasonably well.
The statistical error in the determination of $R_{br}$ after two
years at an $e^+e^-$ linear collider with $\sqrt{s} = 300$ GeV
is 17$\%$.\cite{Nakamura}
We also need to know the theoretical ambiguity of the
calculation of the branching
ratios in $h \rightarrow b\bar{b}, c\bar{c}, gg$.
At the moment the theoretical error in the calculation of $R_{br}$
is estimated to be rather large ($\sim$ 20$\%$) due to
uncertainties in $\alpha_s$ and $m_c$.\cite{Kamoshita,DSZ}
But these uncertainties can be reduced in future from both
theoretical and experimental improvements. 

\section{B and K decay in the supergravity model}  
\subsection{Flavor mixing in the supergravity model}
In the minimal SM various FCNC processes and CP violation 
in B and K decays are determined by the CKM matrix. The constraints
on the parameters in the CKM matrix elements $V_{ij}$ can be 
conveniently expressed
in terms of the unitarity triangle which is based on the unitarity
relation $V_{ud} V_{ub}^*+V_{cd} V_{cb}^*+V_{td} V_{tb}^*=0$ 
in the complex plane. With CP violation at B factory as well as
rare K decay experiments we will be able to check consistency of the 
unitarity triangle and at the same time search for effects of physics 
beyond the SM. In order to distinguish possible new physics effects it 
is important to identify how various models can modify the SM predictions. 

As we discussed in section 2.2 the squark and the slepton
mass matrices become new sources of flavor mixings in the 
SUSY model and generic mass matrices would induce too large FCNC 
and LFV effects if the superpartners' masses are in the 100 GeV
region. In the SUSY model based on the supergravity these flavor problems
can be avoided by setting SUSY breaking mass terms universal at the 
very high energy scale. In fact all the scalar fields are assumed to have 
the same SUSY breaking mass at the Planck scale in the minimal 
supergravity model and therefore there are no FCNC and LFV at this 
scale. Physical squark and slepton masses defined at the
weak scale are determined through RGE. General consequences 
are: (i) Squarks for the first and second generations remain 
highly degenerate so that the constraint from the 
$K^0 - \bar{K}^0$ mixing can be safely satisfied. (ii) Due to 
the effect of large top Yukawa coupling constant the stop
and the sbottom can be significantly lighter than other squarks. 
This will induce sizable contributions to FCNC processes such as 
$b \rightarrow s\gamma$ \cite{BBMR,GO}, $b \rightarrow sl^+l^-$
\cite{BBMR,bsll}, $\Delta M_B$ \cite{BBMR,GNO},
$\epsilon_K$ \cite{GNO} and $K \rightarrow \pi \nu \bar{\nu}$.
(iii) In the SUSY GUT the large top Yukawa coupling constant also 
induces the flavor mixing in the slepton sector so that LFV processes
such as $\mu^+ \rightarrow e^+ \gamma$, 
$\mu^+ \rightarrow e^+e^+e^-$ and $\mu^- - e^- $ conversion in 
atoms receive large SUSY contributions.\cite{BH} 
In this section we deal with the processes listed in 
(ii) following recent update of the calculation \cite{GOS} 
and the LFV processes are discussed in the next section.

In the minimal supergravity model we can introduce four new complex 
phases in SUSY breaking terms and the $\mu$ term, of which only
two are physically independent phases. These phases in general induce 
too large electron and neutron EDMs if the phase is $O(1)$ and the
squark and slepton masses are a few hundred GeV.\cite{EDM} Here we 
assume that all these parameters are real so that the source 
of the CP violation is only in the CKM matrix element. In this case 
we can show by evaluating the RGE that the SUSY loop contributions to FCNC 
amplitudes approximately have the same dependence on the CKM elements 
as the SM contributions. In particular, the complex phase of the 
$B^0 - \bar{B^0}$ mixing amplitude does not change even if we take 
into account the SUSY and the charged Higgs loop contributions. In 
terms of the unitarity triangle this means that the angle measurements 
through CP asymmetry in B decays determine the CKM matrix elements as 
in the SM case. On the other hand the length of the unitarity triangle 
determined from $\Delta M_B$ and $\epsilon_K$ can be modified.

If we consider the supergravity model with SUSY CP phases we have to
take into account the constraints from EDM explicitly. Even in such 
case it is shown that the the phase of  the $B^0 - \bar{B^0}$ mixing 
amplitude cannot be much different from the SM amplitude.\cite{Nihei}

There is important difference between two classes of the processes
in constraining the SUSY parameter space, namely,(i) 
$b \rightarrow s\gamma$ and $b \rightarrow sl^+l^-$ and
(ii) $\Delta M_B$, $\epsilon_K$ and $K \rightarrow \pi \nu \bar{\nu}$.
While the class (i) processes only depend on CKM parameters which
is already well known, the class (ii) processes depend on $V_{td}$
element or $(\rho,\eta)$ in the Wolfenstein parameterization.
Present constraint on $(\rho,\eta)$ which is independent of 
$\Delta M_B$ and $\epsilon_K$ only comes from the  $b\rightarrow
u$ transition, thus there are still large uncertainty.
While the class (i) processes, especially the $b \rightarrow s\gamma$ 
process, already provide important constraint on possible SUSY loop effect,
the class (ii) processes become useful constraint after new information
on the unitarity triangle is obtained at B factories.

\subsection{$b \rightarrow s\gamma$, $b \rightarrow sl^+l^-$}
First we discuss the branching ratios of $b \rightarrow s\gamma$, 
$b \rightarrow sl^+l^-$ in supergravity model.\cite{GOS} 
In the calculation in this and the next subsection we have used updated 
results of various SUSY search experiments at LEP2 and Tevatron
as well as the next-to-leading QCD corrections in the calculation
of the $b \rightarrow s\gamma$ branching ratio. 
We calculated the SUSY particle spectrum based on two different
assumptions on the initial conditions of RGE.  
The minimal case corresponds to the minimal supergravity where 
all scalar fields have a common SUSY breaking mass at the GUT scale. 
In the second case shown as ``all'' in the figures we enlarge the SUSY 
parameter space by relaxing the initial conditions for the SUSY breaking 
parameters, namely all squarks and sleptons have a 
common SUSY breaking mass whereas an independent
SUSY breaking parameter is assigned for Higgs fields. 

The Br($b \rightarrow s\gamma$) and Br($b \rightarrow sl^+l^-$) can be 
calculated using the weak effective Hamiltonian  
\begin{equation}
H_{eff}= \sum_{i=1}^{10} C_i O_i + h.c..
\end{equation}
The relevant operator for the calculation of
$b \rightarrow s\gamma$ process is $O_7$ and
the $b \rightarrow sl^+l^-$ process 
also depends on $O_9$ and $O_{10}$. These operators are 
defined as
\begin{eqnarray}
O_7&=&\frac{e}{16\pi^2}
m_b (\bar{s}_L \sigma_{\mu \nu} b_R) F^{\mu \nu},\\
O_9&=&\frac{e^2}{16\pi^2}
(\bar{s}_L \gamma_\mu b_L)(\bar{l}_L \gamma^\mu l_L),\\
O_{10}&=&\frac{e^2}{16\pi^2}
(\bar{s}_L \gamma_\mu b_L)(\bar{l}_L \gamma^\mu \gamma_5 l_L).
\end{eqnarray}

The effect of SUSY particle and charged Higgs loop can be taken
into account in the calculation of the coefficient functions for
these operators at the weak scale. By numerical calculation
it was shown that only the coefficient of $O_7$ receives sizable 
correction in this model. Moreover, although the charged Higgs boson 
loop contribution has the same sign as the SM one the SUSY loop 
effect can interfere with the SM amplitude either destructively
and constructively. 

In Figure 7 the $b \rightarrow s\gamma$ branching ratio
is shown as a function of charged Higgs mass for $\tan{\beta}=2$. 
The dark points correspond to the minimal case and the light
points to the enlarged parameter space. We can see that unlike
the type II two Higgs doublet model the charged Higgs boson mass 
less than 400 GeV is allowed because of destructive interference.
\begin{figure}
\begin{center}
\mbox{\psfig{figure=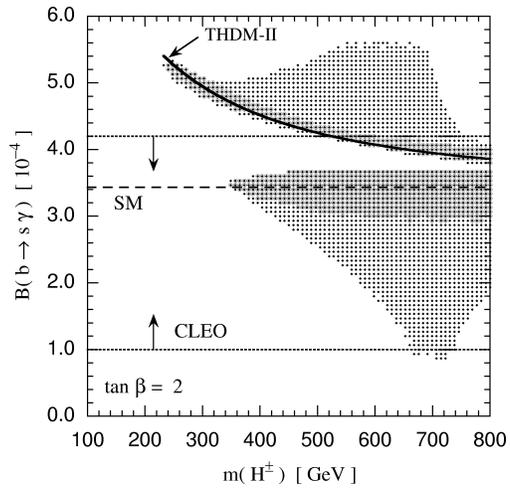,width=2.6in,angle=0}}
\end{center}
\caption{Br($b \rightarrow s\gamma$) in the supergravity
model as a function of the charged Higgs boson mass for
$\tan{\beta}$=2. The solid line shows the branching ratio
in the two Higgs doublet model (Model II). The dark points 
represent the results for the minimal supergravity and the light points
correspond to the extended parameter space where SUSY soft
breaking parameters at the GUT scale are different for
squark/slepton and Higgs fields. Also the SM prediction and
the CLEO 95 bound
\protect\cite{CLEO}
are shown.  
\label{fig:fig7}
}
\end{figure}      
Since only the coefficient of $O_7$ is significantly affected
by the SUSY contribution there is a strong correlation between
the $b \rightarrow s\gamma$ and $b \rightarrow sl^+l^-$
branching ratios. Figure 8 shows this correlation for
$\tan{\beta}=30$. Here $b \rightarrow s\mu^+\mu^-$ branching
ratio is integrated over lepton invariant spectrum below the 
$J/\psi$ threshold to avoid the large resonance peak. We can see that
there are two fold ambiguity which corresponds to the sign of the 
coefficient of $O_7$.
When the SUSY and charged Higgs contribution is (-2) times the SM
contribution the  $b \rightarrow s\gamma$ branching ratio is
the same as the SM while the $b \rightarrow sl^+l^-$ branching
ratio can be enhanced by about a factor of two. In the supergravity model
this situation arises only for large $\tan{\beta}$ case. The lepton
forward and backward asymmetry can also show sizable deviation
from the SM in this case.
\begin{figure}
\begin{center}
\mbox{\psfig{figure=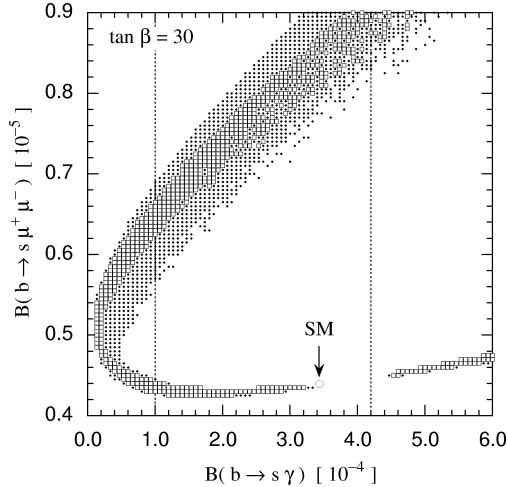,width=2.6in,angle=0}}
\end{center}
\caption{Correlation between Br($b \rightarrow s\gamma$) and 
Br($b \rightarrow sl^+l^-$) in the supergravity
model for $\tan{\beta}$= 30. Here Br($b \rightarrow s\mu^+\mu^-$)
is obtained by integrating in the range $2 m_{\mu} < \sqrt{s} <
m_{J/\psi} - 100$ MeV where $\sqrt{s}$ is the invariant mass of
$\mu^+ \mu^-$ pair. The vertical lines correspond to the CLEO 95 
upper and lower bounds.
\label{fig:fig8}
}
\end{figure}     

\subsection{ $\Delta M_B$, $\epsilon_K$,
$K_L \rightarrow \pi^0 \nu \bar{\nu}$}
We have calculated the $\Delta M_B$, $\epsilon_K$ and 
branching ratios of $K_L \rightarrow \pi^0 \nu \bar{\nu}$
and $K^+ \rightarrow \pi^+ \nu \bar{\nu}$ in the SUSY model based
on supergravity\cite{GOS}. 
In Figures 9 and 10 we present  $\Delta M_{B_d}$ and 
Br($K_L \rightarrow \pi^0 \nu \bar{\nu})$ in the present model 
normalized by the same quantities calculated in the SM as the function
of the $b \rightarrow s\gamma$ branching ratio. Note that these ratios 
are essentially independent of the CKM parameters because the SUSY and 
the charged Higgs boson loop contributions have the same dependence on 
the CKM parameters.  Although we only present the results for 
$\Delta M_{B_d}$ and Br($K_L \rightarrow \pi^0 \nu \bar{\nu})$,
$\epsilon_K$ and Br($K^+ \rightarrow \pi^+ \nu \bar{\nu}$)
provide the same constraints on the SUSY parameters respectively
because these quantities are almost equal if normalized by corresponding
quantities in the SM. 

From Figures 9 and 10 we can conclude that 
the $\Delta M_{B_d}$ (and $\epsilon_K$) can be enhanced by up 
to 40\% and Br($K_L \rightarrow \pi^0 \nu \bar{\nu})$ (and 
Br($K_+ \rightarrow \pi^+ \nu \bar{\nu}$))is suppressed by up 
to 10\% for extended parameter space and the corresponding numbers 
for the minimal case are 20\% and 3\%. The ratio of two
Higgs vacuum expectation value, $\tan{\beta}$, is 2 for these figures and 
the deviation from the SM turns out to be smaller for large value of    
$\tan{\beta}$.
\begin{figure}
\begin{center}
\mbox{\psfig{figure=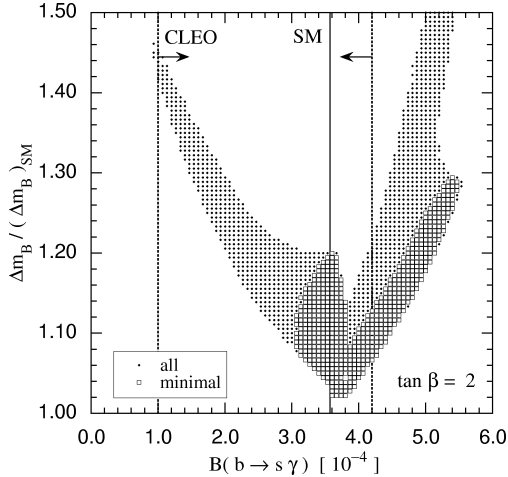,width=2.6in,angle=0}}
\end{center}
\caption{$\Delta M_{B_d}$ normalized by the SM value for $\tan{\beta}=2$
as a function of $b \rightarrow s\gamma$ branching ratio. The dark
(light) points correspond to the minimal (enlarged) parameter space of
the supergravity model. The vertical lines correspond to the CLEO 95 
upper and lower bounds.
\label{fig:fig9}
}
\end{figure} 
\begin{figure} 
\begin{center}
\mbox{\psfig{figure=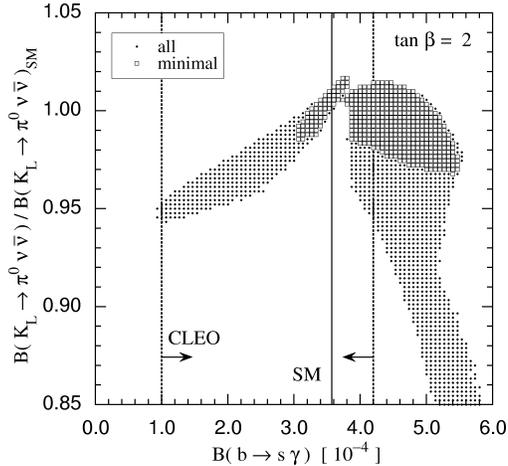,width=2.6in,angle=0}}
\end{center}
\caption{ Br($K_L \rightarrow 
\pi^0 \nu \bar{\nu})$ normalized by the SM value for $\tan{\beta}=2$
as a function of $b \rightarrow s\gamma$ branching ratio.
\label{fig:fig10}}
\end{figure} 

These deviations may be evident in future when B factory experiments 
provide additional information on the CKM parameters. It is expected 
that the one of the three angles of the unitarity triangle is determined 
well through the $B \rightarrow J/\psi K_S$ mode. Then assuming the SM,
one more physical observable can determine the CKM parameters or 
$(\rho,\eta)$ in the Wolfenstein parameterization.
New physics effects
may appear as inconsistency in the determination of these parameters 
from different inputs. For example, the $\rho$ and $\eta$ parameters 
determined from CP asymmetry of B decay in other modes, $\frac
{\Delta M_{B_s}}{\Delta M_{B_d}}$ and $|V_{ub}|$ can be considerably
different from those determined through  $\Delta M_{B_d}$ 
$\epsilon_K$ and Br($K\rightarrow \pi \nu \bar{\nu})$ because
$\Delta M_{B_d}$ $\epsilon_K$ are enhanced and 
Br($K\rightarrow \pi \nu \bar{\nu})$'s are suppressed in the present
model. It turns out that there is strong correlation between 
deviation of two quantities, namely the latters are most
suppressed where the formers have the largest deviation from the SM.
The pattern of these deviations from the SM will be a key 
to distinguish various new physics effects. We also note from Figure
\ref{fig:fig9} and \ref{fig:fig10} that, although the new results 
reported at ICHEP98
($2.0\times 10^{-4}<$Br$(b\rightarrow s\gamma)<4.5\times 10^{-4}$)
\cite{CLEO98} does not change the situation very much,
future improvement on the $b\rightarrow s\gamma$ branching will give
great impacts on constraining the size of possible deviation from 
the SM in FCNC processes. 

\section{LFV in the SUSY GUT }
\subsection{$\mu^+ \rightarrow e^+ \gamma$ in SUSY GUT}
In the minimal SM with massless neutrinos the lepton number is 
conserved separately for each generation. This is not
necessarily true if we consider physics beyond the SM. 
LFV can easily occur in many 
extension of the SM. Experimentally, LFV searches 
are continued in various processes and especially strong
bounds are obtained for muon processes such as
$\mu^+ \rightarrow e^+ \gamma$, $\mu^+ \rightarrow e^+ e^+ e^-$
and $\mu^-  - e^-$ conversion in atoms. The experimental upper bound on 
these processes quoted in PDG 98 are Br($\mu^+ \rightarrow e^+ \gamma$) 
$\leq 4.9 \times 10^{-11}$, Br($\mu^+ \rightarrow e^+ e^+ e^-$)
$\leq1.0 \times 10^{-12}$ and 
$\frac{\sigma(\mu^-T_i \rightarrow e^-T_i)}
{\sigma(\mu^-T_i \rightarrow caputure)}\leq 4.3 \times 10^{-12}$.
Recently there are considerable interests on these processes because
predicted branching ratios turn out to be close to the upper bounds in 
the SUSY GUT.\cite{BH}  

As discussed in previous sections no LFV is generated at the Planck scale
in the context of the minimal supergravity model. In the SUSY GUT scenario,
however, the LFV can be induced through renormalization effects on 
slepton mass matrix because the GUT interaction breaks lepton flavor
conservation. In the minimal SUSY SU(5) GUT, the effect of the large 
top Yukawa coupling constant results in the LFV in the right-handed 
slepton sector. This induces $\mu^- \rightarrow e^-_R \gamma$ 
($\mu^+ \rightarrow e^+_L \gamma$) decay. The branching ratio
can be calculated based on this model and it was pointed out
that there is unfortunate cancellation between different diagrams 
so that  Br($\mu^+ \rightarrow e^+ \gamma$) is below $10^{-13}$ level
for most of the parameter space.\cite{HMTY2} 

In the SO(10) SUSY GUT model, on the other hand, both left- and 
right-handed sleptons induce LFV because of different structure of 
Yukawa coupling constants at the GUT scale.  More importantly,
there are diagrams which enhance the branching ratio by a factor
$(\frac{m_\tau}{m_\mu})^2$ compared to the minimal SU(5) SUSY GUT
model and  predicted branching ratio is at least larger by two
order of magnitudes.\cite{BHS} 

\begin{figure} 
\begin{center}
\mbox{\psfig{figure= 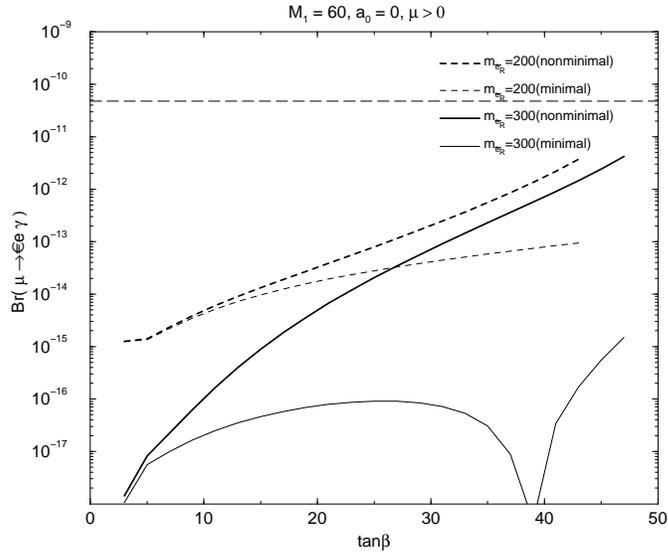,width=4in,angle=-90}}
\end{center}
\caption{
Dependence of the branching ratio of $\mu\rightarrow e \gamma$
on $\tan\beta$ for  the right-handed selectron mass 200GeV 
(dashed lines) and 300GeV (solid lines). The thick lines are for 
the non-minimal case that $V_{\bar{e}}$ and $V_{l}$ are the same as 
$V_{\rm KM}$, and the thin lines are for the minimal case in which 
$V_{\bar{e}}=V_{\rm KM}$ and $V_{l}={\bf 1}$. In this figure we 
choose the bino mass 60GeV, the coefficient of the A term 
$a_0$ =0, the higgsino mass positive.  The long-dashed line is 
the experimental upper bound.
\label{fig:fig11}}
\end{figure}

A similar enhancement can occur 
also in the context of the SUSY SU(5) model for large value of
$\tan{\beta}$ once we take into account effects of higher dimensional 
operators to explain realistic fermion masses.\cite{HNOST} 
In the minimal case the Yukawa coupling is given by 
the superpotential $W= (y_u)_{ij}{\bf T_i}\cdot {\bf T_j}\cdot {\bf H(5)}
+(y_d)_{ij}{\bf T_i}\cdot{\bf \bar{F}_j}\cdot {\bf \bar{H}(5)}$
where ${\bf T_i}$ is 10 dimensional and ${\bf \bar{F}_j}$
is 5 dimensional representation of SU(5). 
This superpotential alone cannot explain the lepton and quark 
mass ratios for the first and second generations although
the $m_b/m_{\tau}$ ratio is in reasonable agreement.
One way to obtain realistic mass ratios are to introduce
higher dimensional operators such as $\frac{f_{ij}}{M_{Planck}}
{\bf \Sigma(24)}\cdot{\bf T_i}\cdot{\bf \bar{F}_j}
\cdot{\bf \bar{H}(5)}$. We investigated how
inclusion of these terms changes prediction of the branching ratio.  
It turns out that the branching ratio is quite sensitive to the
details of these higher dimensional operators. Firstly, once we
include these terms the slepton mixing matrix elements $\lambda_
{\tau}\equiv V_{\tilde{e}31}^*V_{\tilde{e}32}$ which appear
in the formula of the $\mu^+ \rightarrow e^+ \gamma$
amplitude is no longer related to the corresponding CKM matrix 
elements. More importantly, for large value of $\tan{\beta}$,
the left-handed slepton also induces the LFV and the predicted branching 
ratio becomes enhanced by two order of magnitudes as in the SO(10) 
case.\cite{ACH} The destructive interference among the different diagrams
also disappear. We show one example of such calculation in Figure 
\ref{fig:fig11} where Br($\mu^+ \rightarrow e^+ \gamma$) can be close 
to $10^{-11}$ level for large values of $\tan{\beta}$ in the 
non-minimal case.

Another example of possible large LFV is a SUSY model with the 
see-saw type neutrino mass.\cite{Rneutrino} If we include the 
right-handed neutrino supermultiplet at high energy scale, we can include
large Yukawa coupling constant among lepton doublet, right-handed
neutrino and Higgs doublet field. The renormalization effect
can induce the LFV in the left-handed slepton sector and in this 
case $\mu^- \rightarrow e^-_L \gamma$ 
($\mu^+ \rightarrow e^+_R \gamma$) decay can occur.
The branching ratio depends on, among other things, the
magnitude of the Yukawa coupling constant
which is a function of the neutrino mass and the Majorana mass scale.
Therefore even if we fix the neutrino mass motivated by atmospheric
and solar neutrino problem  the branching ratio strongly depends on
the right-handed neutrino mass scale. 
If the Yukawa coupling constant is assumed to be as large as the top 
Yukawa coupling constant the predicted the branching ratio can 
reach the present experimental upper bound. Note that without SUSY 
neutrino mass can only induce negligible effect to the LFV processes
for $\mu$ and $\tau$ decay. It is therefore interesting that the same 
source of LFV can be responsible for the neutrino oscillation and 
the $\mu \rightarrow e \gamma$ decay in the SUSY model.

\subsection{LFV process with polarized muons}
Finally we would like to discuss usefulness of the polarized 
muons in search for LFV.  A highly polarized muon beam is available in 
$\mu^+$ decay experiments. Muons from $\pi^+$ decay stopped near 
the surface of pion production target is 100\% polarized opposite
to the muon momentum and this muon is called surface muon.

The first obvious merit of polarized muons in 
$\mu^+ \rightarrow e^+ \gamma$ is that we can distinguish   
$\mu^+ \rightarrow e^+_R \gamma$ and 
$\mu^+ \rightarrow e^+_L \gamma$ by the angular distribution
of the decay products with respect to the muon polarization direction.
For example, the positron from the $\mu^+ \rightarrow e^+_R \gamma$
decay follows the $(1-P \cos{\theta})$ distribution where $\theta$ 
is the angle between the polarization direction and the positron
momentum and $P$ is the the muon polarization. 
In the previous examples the SU(5) SUSY GUT for small $\tan{\beta}$
predicts $\mu^+ \rightarrow e^+_L \gamma$
because LFV is induced only in the right-handed slepton sector.
On the other hand the SO(10) SUSY GUT generates both 
$\mu^+ \rightarrow e^+_L \gamma$ and $\mu^+ \rightarrow e^+_R \gamma$.
If LFV is induced by the right-handed neutrino Yukawa coupling constant,
only $\mu^+ \rightarrow e^+_R \gamma$ should be observed.

Polarized muons are also useful to suppress background processes 
for the $\mu^+ \rightarrow e^+ \gamma$ search.\cite{polmu} 
In this mode the experimental sensitivity is limited by 
appearance of the background processes. There are two major background
processes. The first one is physics background process which is a tail 
of radiative muon decay. If neutrino pair carries out only little energy
in the $\mu^+ \rightarrow e^+ \nu \bar{\nu} \gamma$ process, we cannot
distinguish this from the signal process. The second background process 
is an accidental background process where detections of 52.8 MeV positron
and 52.8 MeV photon from different muon decays coincide within time
and angular resolutions for selection of signals. The source of the  
52.8 MeV positron is the ordinary $\mu^+ \rightarrow e^+ \nu \bar{\nu}$
decay whereas the 52.8 MeV photon  mainly comes from a tail of
the radiative muon decay. In the experiment with polarized muons 
angular distribution is useful to suppress both backgrounds.
Suppression of the accidental background turns out to be more 
important in the next generation experiment.
In this case both positrons and photons 
follow the $(1+P \cos{\theta})$ distribution. Since the signal
is back-to-back the background suppression works independently of 
the signal distribution. If we use 97\% polarized muons we can expect 
to reduce the accidental background by one order of magnitude. This looks 
promising for search of $\mu^+ \rightarrow e^+ \gamma$ at the level 
of $10^{-14}$ branching ratio.
\begin{figure} 
\begin{center}
\mbox{\psfig{figure= 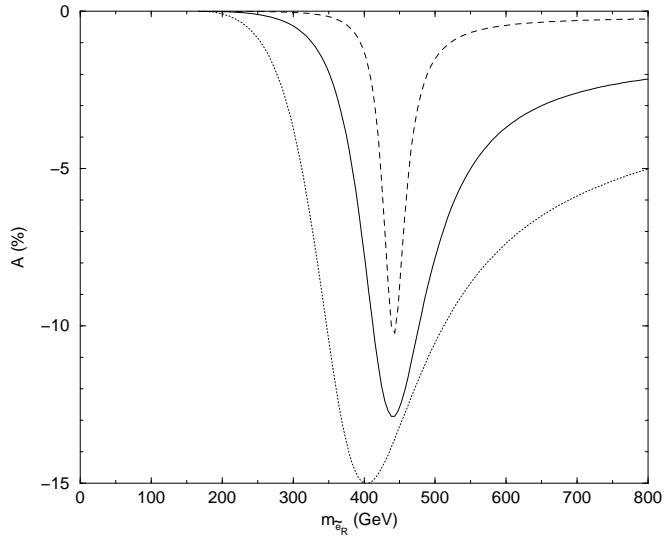,width=4in,angle=-90}}
\end{center}
\caption{~T-odd asymmetry ${\bf A}$ as a function of 
      the right-handed selectron mass in $\mu \rightarrow 3e $ precess 
      in the SUSY SU(5) GUT. We fix the SUSY parameters as 
      $M_2 = 200$ GeV, $A_X = i$, $\mu > 0$ and $\tan\beta = 3$ (dotted
      line), 10 (solid line), 30 (dashed line) and top quark mass
      as 175 GeV.
\label{fig:fig12}}
\end{figure}
 
Last example of usefulness of polarized muons is that we can
measure T violating asymmetry in the $\mu^+ \rightarrow e^+ e^+ e^-$ decay. 
Using polarization of initial muons we can define T odd triple vector 
correlation $<\vec{P}\cdot(\vec{p_1}\times\vec{p_2})>$ where
$\vec{P}$ is muon polarization and  $\vec{p_1}$ and
$\vec{p_1}$ are two independent momenta of decay particles.\cite{m3eCP} 
We have investigated T odd asymmetry in the 
SU(5) SUSY GUT.\cite{OOS} In order to have this asymmetry we need 
to introduce a CP violating phase other than the KM phase. 
This phase can be provided by the complex phases 
in the SUSY breaking terms, for example, the phase
in the triple scalar coupling constant (A term). Since this phase
also induces electron and neutron EDMs, we have calculates the
T odd asymmetry in the $\mu^+ \rightarrow e^+ e^+ e^-$ taking
into account EDM constraints. In Figure\ref{fig:fig12} we show 
one example of such calculation for the T odd asymmetry. By examining 
SUSY parameter space we found that the asymmetry up to 20\% 
is possible. The branching ratio for $\mu^+ \rightarrow e^+ e^+ e^-$ 
turns out to crucially depend on the slepton mixing element 
$\lambda_{\tau}$ which is an unknown parameter once we take 
into account the higher dimensional operators for the Yukawa 
coupling constants. 
For $\lambda_{\tau}=10^{-2}$ we can show that 
the branching ratio of $10^{-14}$ is possible with 10\%
asymmetry which can be reached in future experiment with sensitivity
of $10^{-16}$ level.

\section{Conclusions}
I have reviewed some aspects of the Higgs physics and flavor physics
in SUSY theories. An important feature of SUSY models is
existence of light Higgs boson which is a target of LEP II, Tevatron and
LHC experiments. I have also shown that a future $e^+ e^-$ linear 
collider is an ideal place to study the SUSY Higgs sector.
At earlier stage of the experiment with
$\sqrt{s}\sim$ 300 - 500 GeV, it is easy to find a light
Higgs boson predicted in SUSY standard models.
In particular, both in the MSSM and the SUSY SM with
a gauge singlet Higgs, at least one of neutral Higgs bosons
is detectable. More importantly,
detailed study on properties of the Higgs boson is possible
at an $e^+ e^-$ linear collider through measurements of various 
production cross-sections and branching ratios. As an example 
we show that the measurement of Higgs couplings 
to $c\bar{c}$/$gg$/$b\bar{b}$ gives us important information 
on the Higgs sector of the MSSM.  
Combining expected results at LHC we may be able to clarify the 
Higgs sector of the SM and explore physics beyond the SM such as 
the SUSY SM in future $e^+ e^-$ linear collider experiments.

We have considered various flavor changing processes in the 
supersymmetric standard model based on the supergravity.
We have seen that the branching ratio of $b \rightarrow s\gamma$
is very sensitive to SUSY effects and further improved measurement
of this process is important. 
Flavor changing neutral current processes in B and K decays
such as $B^0 - \bar{B^0}$ mixing, $\epsilon_K$ and branching 
ratio of $K \rightarrow \pi \nu \bar{\nu}$ are calculated and 
it is shown that the deviation from the SM becomes as large as 40 \%
for $B^0 - \bar{B^0}$ mixing and $\epsilon_K$ but somewhat smaller
for $K \rightarrow \pi \nu \bar{\nu}$ processes.
We also investigated LFV in the SU(5) SUSY GUT. 
It is pointed out that the $\mu \rightarrow e
\gamma$ branching ratio can be enhanced for large $\tan{\beta}$
if we take into account the higher dimensional operators
in the Yukawa coupling constants at the GUT scale.
The T odd triple vector correlation is also calculated for the
$\mu^+ \rightarrow e^+ e^+ e^-$ process and it is shown that
the asymmetry up to 20\% is possible due to the CP violating
phases in the supersymmetry breaking terms. 
Experiments on B, K and LFV
processes in near future, therefore, will provide very important
opportunities to investigate into the structure of the SUSY 
breaking sector.


\end{document}